\documentclass[11pt]{article}
\usepackage{amssymb}
\usepackage{pst-all}
\usepackage{pstricks}
\usepackage{pstcol,pst-fill,pst-grad}
\usepackage{amsfonts}
\usepackage{graphicx}
\usepackage{amsmath}
\usepackage[normalem]{ulem}
\usepackage{multicol}
\usepackage{tikz}
\usepackage{enumerate}

\usetikzlibrary{matrix}
\RequirePackage{mathrsfs} \RequirePackage[sc]{mathpazo}
\RequirePackage{wasysym} \RequirePackage{setspace}
\makeatletter \@addtoreset{equation}{section}

\newcommand{\be}{\begin{equation}}
\newcommand{\ee}{\end{equation}}
\newcommand{\bea}{\begin{eqnarray}}
\newcommand{\eea}{\end{eqnarray}}
\input{tcilatex}
\begin{document}

\date{}
\title{Axions in Type IIA Superstring Theory and Dark Matter}
\author{ Adil Belhaj$^{a}$ \thanks{%
belhaj@unizar.es}, Salah-Eddine Ennadifi$^{b}$ \thanks{%
ennadifis@gmail.com}, \\
\\
{\small $^{a}$ LIRST, Polydisciplinary Faculty, Sultan Moulay Slimane
University }\\
{\small B\'eni Mellal, Morocco } \\
{\small $^{b}$ LHEP-MS, Faculty of Science, Mohammed V University, Rabat,
Morocco} }
\maketitle

\begin{abstract}
Motivated by the M-theory compactification physics, we study a type
IIA superstring quiver model with  an open string axion field
associated with the U(1)$_{PQ}$ Peccei-Quinn global symmetry using
D6-brane physics. This realization extends the symmetry of the
SM-Like and contributes to the mediation of the electroweak symmetry
breaking. The associated    axion energy scales are approached in
terms of D2 and D6-branes in type IIA compactifications on
Calabi-Yau threeforlds. Using the fermion mass scales as well as
string theory results, the allowed axion window is discussed in
terms of intersecting D6-branes wrapping 3-cycles belonging to the
middle cohomology of the type IIA internal space. The
non-perturbative effect of the open string axion sector is dealt
with using a D2-brane wrapping a 3-cycle producing (1+3)-dimensional
stringy instantons. According to the known data, the vacuum
expectation value of the axion and the mass scale of the involved
scalar provide a probe of the stringy physics effect in the SM and a
possible axionic dark matter candidate.

\textbf{Keywords}: Type IIA superstring,  D-branes, Calabi-Yau
spaces, Open string  axions, Dark Matter.
\end{abstract}

\newpage

\section{Introduction}

It has been shown that the physics of low-energy phenomena is successfully
described by the so-called Standard Model (SM) of particle physics. With the
recent observations including the one associated with the Higgs boson, a
large amount of experimental data is in a good agreement with SM physics\cite%
{1,2,3}. However, certain famous issues mostly related to the
hierarchy problem, fermion masses and Dark Matter (DM) are still
open scientific subjects \cite{4,5}. To deal with such questions,
models beyond SM have been built using theories based on extra
dimensions of spacetime and supersymmetry in the context of
string/M-theory. In this way, the extra dimensions should be compact
in such models producing (3+1)-dimensional physics with the presence
of massless pseudo scalar axion like fields. To examine the physics
of such fields, many stringy inspired models have been proposed
using either geometric engineering method in M-theory on G2
manifolds or intersecting type II D-branes wrapping non trivial
cycles embedded in complicated geometries including orbifold
Calabi-Yau manifolds [6-26]. In particular, interesting type II
superstring  models in the presence of various types of axionic
fields have been studied using different ways. It has been shown
that such fields can play a primordial r\^{o}le in the physics
beyond the SM and cosmology \cite{17,18}. Indeed, globally
consistent D6-brane models with SM spectrum in the type IIA
orientifold compactification scenario have been constructed.
Moreover,
different energy scales have been examined including the Peccei-Quinn U(1)$%
_{PQ}$ symmetry corresponding to the electroweak and the supersymmetry
breaking scales in the context of closed and open string theory sectors
[10-16].

The axion fields are characterized by two parameters: the axion
decay constant and an energy scale derived from non-perturbative
effect. This effect is associated with instantons in superstring
theory. Moreover, it has been remarked that the number of such
axions depends on the topology of the corresponding internal
geometries in string/M-theory. More recently, DM axions in such
theories have been investigated using hidden particle
sectors\cite{260,261}. In particular, a study from M-theory
compactified on G2-manifolds has been conducted. Using string
duality, this could be related to intersecting numbers of 3-cycles
in Calabi-Yau manifolds in the presence of D6-branes.

The objective of this paper is to contribute to these investigations
by studying  type IIA superstring quivers with an open string
axion field associated with the U(1)$_{PQ}$ Peccei-Quinn global
symmetry. In the case of standard singularities, this realization
extends the symmetry of the SM-Like and contributes to the mediation
of the electroweak symmetry breaking. The associated, open string,
axion energy scales are discussed in terms of D2 and D6-branes in
type IIA compactification. Using the fermion mass scales as well as
string theory data, the allowed axion window is treated in terms of
intersecting D6-branes wrapping on 3-cycles belonging to the middle
cohomology of the type IIA internal space. However, the non
perturbative effect in the open string axion field is approached
using a D2-brane wrapping on a 3-cycle producing stringy instantons
in 1+3 dimensions. According to the known results, the vacuum
expectation value of the axion and the mass scale of the involved
new scalar provide a probe of the stringy physics effect in the SM
and a possible axionic DM candidate.

\section{D-brane model and field content}

In this section, we elaborate a stringy axion model using string
theory approach. It has been argued that open strings stretched to
different stacks of type II D-branes generate the SM particle
spectrum. More precisely, an unitary group $\mbox{U}(N)\sim
\mbox{SU}(N)\times \mbox{U}(1)$ is obtained by considering $N$
coincident type II D-branes. Typically, this gauge group model can
be extended by introducing extra abelian factors \cite{9,10}.
Generically, these abelian fields involve (1+3)-dimensional
anomalies canceled via the Green-Schwarz mechanism generating a mass
to the anomalous U(1) fields by breaking the associated gauge
symmetry \cite{10}. This gives an acceptable effective low-energy
description reproducing either SM-like or more generally some
possible extensions \cite{20,21}. In such models, the gauge and the
matter fields could be encoded in quivers representing stacks of
coinciding D-branes in type II superstrings. Working at the level of
quivers one might investigate the Yukawa-like couplings as well as
their strengths from the quiver data rather than concentrating on
the whole geometrical and physical data derived from string theory.
Here, we build a type IIA quiver model involving an axion-like
particle using the intersecting D-brane method. Type IIA D-brane
configurations with orientifold geometries have been considered to
produce non trivial components associated with SM-like theories. To
get this model, we consider a set of three stacks of D6-branes which
can accommodate the U(3)$\times $ Sp(1) $\times $U(1) gauge symmetry
and SM matter fields. A way to deal with the axion field is to use
the U(1)$_{PQ}$ symmetry. It has been observed that there are many
roads to handle such a symmetry considered either as a local or as a
global one depending on the model in question. To consider
this symmetry, we can enlarge the $\mbox{U}(3)\times \mbox{Sp}(1) \times %
\mbox{U}(1)$ gauge symmetry by thinking about an extra abelian gauge
factor corresponding to the U(1)$_{PQ}$ symmetry. However, this
symmetry will be viewed as as a global one. Precisely, the present
proposed system will be constructed form four blocks of intersecting
type IIA D6-branes in the presence of a flavor-dependent symmetry.
It has been shown that the symmetry distinguishes various fermions
from each other. In  type IIA  superstring theory, the intersecting
D6-brane configurations produce the gauge group
\begin{equation}
\mbox{U(3)}_{a}\times \mbox{Sp(1)}_{b}\times \mbox{U(1)}_{c}\times %
\mbox{U(1) }_{PQ}.  \label{eq1}
\end{equation}%
In this D-brane industry, the intersections between such D6-branes
generate matter fields. Investigations reveal that the associated
physics can be encoded nicely in a quiver diagram, considered as a
possible generalization of Dynking diagrams of Lie symmetries. The
U(1)$_{PQ}$ symmetry will be broken by incorporating a complex
scalar field $\phi =\rho exp(\frac{i\sigma }{f_{\sigma }})$ where
$f_{\sigma }$ denotes the axion decay constant. Here, the field
$\sigma $ will be considered as an open  string axion obtained from
type IIA superstring using intersecting D6-brane associated with the
U(1)$_{PQ}$ symmetry. It is noted that $\rho$ can be associated with
fluctuations of the corresponding  open string sector.  The complex
scalar $\phi$ gets a non zero vacuum expectation value $\langle \phi
\rangle =f_{\sigma }$. It has been shown that the lagrangian
involving this open  string axion field $\sigma $ has the following
terms
\begin{equation}
\mathcal{L}_\sigma \simeq \partial _{\mu }\sigma \partial ^{\mu }\sigma +\frac{\sigma }{%
f_{\sigma }}F_{\mu\nu}\wedge F^{\mu\nu},  \label{eq4}
\end{equation}
where $\sigma $ is the phase of the complex scalar  field $\phi$
charged under  the U(1)$_{PQ}$ symmetry. In what follows,
$F_{\mu\nu}$ will be associated with the corresponding abelain gauge
field strength obtained from the quantization of the open string
living on the D6-brane world volume. It is noted that the vanishing
of the non-abelian anomalies is related to the tadpole conditions.
However, the Green-Schwarz mechanism kills the abelian and mixed
anomalies. On the perturbative level, the anomalous U(1)$^{\prime
}s$ become massive and can survive only if they are considered as
global symmetries which cancel various couplings in the theory in
question. In the corresponding gauge symmetry, the hypercharge can
be identified with a massless remained linear combination of the
global symmetries.  The SM fermion charges can be obtained by using
the vanishing of anomaly constraints imposed by the flavor
conditions associated with the complex scalar field $\phi $. In 1+3
dimensions, the complex field gives allowed invariant coupling
terms. Concretely, there are many ways to implement such interaction
coupling terms. It is possible to write down a general form which
can be embedded in SM-like
\begin{equation}
y_{f}\left( \frac{\phi }{M_{s}}\right) ^{n_{f}}hf\overline{f}.  \label{eq6}
\end{equation}%
In this equation, $f$ denotes the SM fermion fields and the coupling
parameters $y_{f}$ indicate the Yukawa constants. $M_{s}$ is the string
scale mass while $n_{f}$ is a number which will be fixed later by
considering SM and string theory known data. The global U(1)$_{PQ}$ symmetry
which acts on the fields as follows
\begin{equation}
\phi \rightarrow e^{iq_{\phi }\alpha }\phi ,\quad h\rightarrow
e^{iq_{h}\alpha }h,\quad f\rightarrow e^{iq_{f}\alpha }f,  \label{eq7}
\end{equation}%
can be used to establish constraints on $n_{f}$ and th gauge charges. The
invariance of the equation (\ref{eq6}) gives the expression of $n_{f}$
\begin{equation}
n_{f}=-\frac{q_{h}+q_{f}+q_{\overline{f}}}{q_{\phi }}.  \label{eq8}
\end{equation}%
Using the SM fields, this equation can be solved using a possible physical
configuration based on intersecting D6-brane charge assignments of type IIA
superstring compactification. In the present work, we propose a particular
system obtained from the intersecting 3-cycles belonging to the middle
cohomology of the internal space. This model can be supported by the known
result of intersecting D6-brane model building using the toroidal
orientifold compactification. To get the above stringy gauge symmetry, we
consider four stacks of D6-branes indicated by D$6_{a,b,c,PQ}$-branes. In
this way, the intersecting numbers $I_{\alpha \beta }$ $=\Sigma _{\alpha
}\circ \Sigma _{\beta }$ between the D$6_{\alpha =a,b,c,PQ}$-brane and D$%
6_{\beta =a,b,c,PQ}$-brane can be given in terms of the 3-cycles $\Sigma
_{\alpha ,\beta }$ intersections. It has been remarked that type IIA
superstring theory compactified on a factorized six dimensional tori $%
T^{6}=T_{1}^{2}\times T_{2}^{2}\times T_{3}^{2}$ can be exploited to compute
the number $I_{\alpha \beta }$ by exploring winding numbers. The model we
study here requires D6-brane configurations with the following intersection
numbers
\begin{eqnarray}
I_{D_{a}D_{b}} &=&3,\quad I_{D_{a}D_{c}}=-2,\quad I_{D_{a}D_{c^{\ast }}}=-1,
\notag \\
I_{D_{a}D_{PQ}} &=&-1,\quad I_{D_{a}D_{PQ^{\ast }}}=-2,\quad
I_{D_{PQ}D_{b}}=3,  \label{eq9} \\
I_{D_{PQ}D_{c}^{\ast }} &=&-3.  \notag
\end{eqnarray}%
These intersection numbers can produce the associated spectrum. In
particular, the gauge fields, the leptons, the the quarks and scalar fields
can be represented by a non trivial graph called a quiver as illustrated in
figure 1.

 \begin{figure}
\begin{center} \includegraphics[width=9cm]{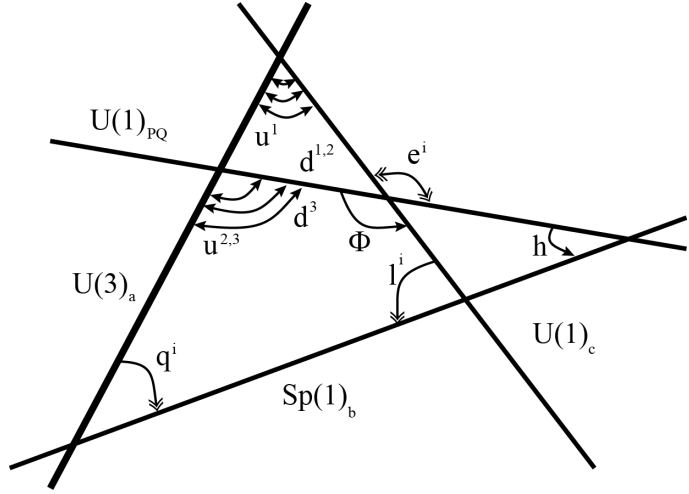}
\caption{Graph of the intersecting D6-brane model.}
\end{center}
\end{figure}

In this graph, the arrows represent the gauge fields and the nodes
indicate the matter fields with the arrows indicate their
fundamental (antifundamental) representations. This stringy system,
which is based on the above intersecting D6-branes, is determined by
a particular choice of the charges which are listed in the table 1.

\begin{center}
\begin{tabular}{|c|c|c|c|c|c|c|c|c|c|}
\hline
Fields & $q_{L}^{i}$ & $\overline{u}_{R}^{1}$ & $\overline{d}_{R}^{1,2}$ & $%
\overline{u}_{R}^{2,3}$ & $\overline{d}_{R}^{3}$ & $l_{L}^{i}$ & $\overline{e%
}_{R}^{i}$ & $h$ & $\phi$ \\ \hline
$\mbox{U}(1)_{c}$ & 0 & 1 & -1 & 0 & 0 & 1 & -1 & 0 & - 1 \\ \hline
$\mbox{U}(1)_{PQ}$ & 0 & 0 & 0 & 1 & -1 & 0 & -1 & 1 & -1 \\ \hline
$\mbox{U}(1)_{Y}$ & 1/6 & -2/3 & 1/3 & -2/3 & 1/3 & -1/2 & 1 & -1/2 & 0 \\
\hline
\end{tabular}

\bigskip Table1: field content corresponding to the Hypercharge combination $%
Y=\frac{1}{6}Q_{a}-\frac{1}{2}Q_{c}-\frac{1}{2}Q_{PQ}$. The index $i(=1,2,3)$
denotes the family index.
\end{center}

\section{Axion and coupling scalar-matter terms}

In this section, we discuss the coupling scalar-matter terms of the
above proposed model. These terms will alow us to get the  open
string axion window appearing in the axion mass expression. Indeed,
the fermion fields are distinguished by their ${\mbox{U}(1)}_{c,PQ}$
charges. This symmetry usually forbids the renormalizable Yukawa
couplings for the ${\mbox{U}(1)}_{PQ}$ charged quarks, but it would
allow them via certain effective highly suppressed couplings
through the complex field $\phi $ carrying an appropriate charge under ${%
\mbox{U}(1)}_{PQ}$. In this set-up, this can be done upon the
simultaneous spontaneous breaking of the Peccei-Quinn and
electro-weak symmetry by the vevs of $\phi$ and $v_{EM}$
respectively. We will see later that string theory requiters
positive numbers for $n_f$. Indeed, the effective Yukawa couplings
appearing in (\ref{eq6}) can be obtained by considering $n_{f}$
solving the charge equations given in (\ref{eq8}) .

The intersection numbers can be used to determine the low-energy effective
Yukawa lagrangian. It has been revealed that the allowed couplings take the
following form
\begin{eqnarray}
\zeta _{Yuk} &=&y_{u^{2}}h^{\dagger }q_{L}\overline{u}_{R}^{2}+y_{u^{3}}h^{%
\dagger }q_{L}\overline{u}_{R}^{3}+y_{d^{3}}hq_{L}\overline{d}_{R}^{3}+\
y_{e^{i}}hl_{L}^{i}\overline{e}_{R}^{i}+y_{u^{1}}\frac{\phi }{M_{s}}%
h^{\dagger }q\overline{u}_{R}^{1}  \notag \\
&+&y_{d^{1}}\frac{\phi ^{\dagger }}{M_{s}}hq\overline{d}_{R}^{1}+y_{d^{2}}%
\frac{\phi ^{\dagger }}{M_{s}}hq\overline{d}_{R}^{2}+y_{v^{i}}\frac{\phi ^{2}%
}{M_{s}^{3}}h^{\dagger }h^{\dagger }l_{L}^{i}l_{L}^{i}+h.c.  \label{eq10}
\end{eqnarray}
The heavy fermions, which are directly related to the electroweak
symmetry breaking sector, have (1+3)-dimensional Yukawa couplings.
However, the lighter ones are associated with the electroweak
symmetry breaking sector via the complex field $\phi $. They have
higher dimensional suppressed
Yukawa-like couplings involving powers of $\left( \phi /M_{s}\right) $ or $%
\left( \phi ^{\dagger }/M_{s}\right) $. The corresponding powers are
required by flavor charges and certain SM physical data. Using the
approximation $\phi ^{2}h/M_{s}^{3}\sim \left( \phi /M_{s}\right) ^{3}$, $%
n_{f}$ should take $0,1$ and $3$. It is observed that $n_{f}=0$
corresponds to the absence of the  open  string axion field. Deeper
investigations of the stringy origin of the axion field can be
explored to give more information on $n_{f}$.

The values of $n_{f}$ will be used to investigate the electroweak symmetry
breaking associated with the lagrangian (\ref{eq10}). After such a breaking
symmetry with $\left\langle h\right\rangle =v_{EW}$, and using a simple
scaling, the resulting lagrangian for the mass coupling term takes the form
\begin{equation}
\zeta _{mass}=y_{f}\epsilon v_{EW}f\overline{f}.  \label{eq13}
\end{equation}

If we forget about the fermion mass hierarchy which could be encoded in the
corresponding Yukawa constants $y_{f}$, $\epsilon =\left( \frac{%
\left
\langle \phi \right \rangle }{M_{s}}\right) ^{n_{f}}$ is a suppression
factor related to the open  string axion decay constant $f_{\sigma}$ and the $%
M_{s}$ via the relation
\begin{equation}
\epsilon ^{\frac{1}{n_{f}}}= \frac{f_{\sigma }}{M_{s}}.  \label{eq14}
\end{equation}
This shows that the decay constant of the open  string axion is
proportional to the mass string scale. Breaking the electroweak
symmetry at the usual scale $v_{EW}\simeq 10^{2}GeV$, and using the
ratio of certain suitable fermion masses, we can estimate the
suppression factor $\epsilon $. In fact, it could be approached by
strange and top quark masses report. It is given by
\begin{equation}
\epsilon \simeq \frac{m_{s}}{m_{t}}\simeq 10^{-3},  \label{eq15}
\end{equation}%
where $m_{s}$ and $m_{t}$ are the strange and the top quark masses.

Taking $M_{s}=10^{12}GeV$, the so-called  open string axion window
can be obtained for the above mentioned values of $n_{f}$. Using the
above data, one can find
the associated energy scales. For $n_{f}=1$, we obtain $f_{\sigma}= 10^{9} %
\mbox{GeV}$ while we get $10^{11}$ for $n_{f}=3$. This is in a good argument
with the astrophysical observations and the cosmological considerations made
on the axion decay constant region \cite{22,23}
\begin{equation}
10^{9}GeV\leq f_{\sigma }\leq 10^{12}GeV.  \label{eq16}
\end{equation}

\section{Axion mass spectrum and dark matter}

Despite the axion directly results from the ${\ \mbox{U(1)}}_{PQ}$ breaking
at the scale $\langle \phi \rangle =f_{\sigma }$ as a solution to the strong
CP problem \cite{31}. It has also been considered as a significant DM
candidate \cite{32}. It naturally meets the dark matter particle
requirements, i.e. cold even though they are so light, non-baryonic and with
extremely weak coupling to ordinary matter as being inversely proportional
to its decay constant (\ref{eq16}). Thus, the only significant long-range
interactions are gravitational. In spite of the fact that the exact mass of
the axion is not known, its mass has been constrained by cosmic observation
and particle physics experiments \cite{33,34,35}. The mass of the axion is
predicted such as
\begin{equation}
m_{\sigma }\simeq 10^{-6}eV\left( \frac{10^{12}GeV}{f_{\sigma }}\right) .
\label{eq17}
\end{equation}%
According to the range of the decay constant in our quiver based on
a D6-brane model, it corresponds to the open string axion mass
\begin{equation}
10^{-2}meV\lesssim m_{\sigma }\lesssim 1meV.
\end{equation}%
This is a mass range for a small elementary particles so that the previous
detectors did not have enough sensibility to coin them. A direct way of the
detection of this kind of particles may be directly by studying their
interaction within the detector. This corresponds to the tiny shocks with
its atomic nuclei, even if this path seems to be more difficult, if not
impossible, due to the axion particle characteristics. Let us make an
estimate of such direct detection. DM axions are supposed to be flying
around the halo of our galaxy with typical speed $v_{\sigma }=10^{-3}c$ \cite%
{36}, and because they are only very weakly interacting, they can go through
everything, just like neutrinos with little trouble. The corresponding
maximal kinetic energy for $m_{\sigma }\simeq 1meV$ reads
\begin{equation}
K_{\sigma }\simeq \frac{1}{2}10^{-6}eV\left( \frac{10^{12}GeV}{f_{\sigma }}%
\right) v_{\sigma }^{2}\sim 10^{-3}\mu eV.  \label{eq18}
\end{equation}%
This value scatters off an atomic nuclei. The energy deposit is only
a highly small recoiling energy; and most of this order of
magnitude. It is extremely a tiny energy deposit that is very
difficult to pick out against background from natural radioactivity
which typically $\sim MeV$. Therefore, a detector with very low
energy threshold will be needed for the hunt for such a  dark matter
target.

In D-brane models, two important physical processes can describe the axion
mass scale as well as its behaviour. In fact, after the breaking of the $%
\mbox{U(1)}_{PQ}$ symmetry at some high scale $f_{\sigma }$ establishing the
axion as a massless Goldstone boson, the non-perturbative physics, in terms
of instantons, obtined from D-branes wrapped on cycles, becomes relevant at
some scale $\Lambda _{\sigma }\ll f_{\sigma }$ and provides a potential for
the axion and thus a tiny mass. The axions then obtain mass by
non-perturbative effects, such as instantons, and are extremely generic
prediction of string theory at low energies. We know read easily that due to
the the large scale hierarchyies $\Lambda _{\sigma }\ll f_{\sigma }\lesssim
M_{s}$, the axion mass is thus parametrically small. Because the scale of
the induced axion potential energy depends exponentially on details of the
internal geometry, and large hierarchies between the non-perturbative scale $%
\Lambda _{\sigma }$ and the string scale $M_{s}$ can easily be achieved.
Explicitly, we have
\begin{equation}
\Lambda _{\sigma }\sim {\mu }e^{-S},
\end{equation}%
where ${\mu }$ is the hard non-perturbative scale, i.e, GUT scale,
SUSY breaking. This could be associated with fermionic zero modes
obtained from open strings living on some branes. $S$ is a instanton
string moduli action describing, in string units, the size of the
3-cycle $\Sigma $ on which the relevant euclidian D-branes are
wrapped. In type II superstring theory, the non-perturbative effects
are principally produced from Euclidean D$p$-branes wrapping
$(p+1)$-cycles \cite{360,361}. The Calabi-Yau compactification
topology, in type IIA superstring, requires that the relevant
instantons are D2-branes wrapped on three-cycles intersecting with
the 3-cycle associated with the D$6_{PQ}$-brane.  In this way, there
are D2-D2 and D2-D$6_{PQ}$ open string sectors producing the
instanton zero modes over which one should  integrate. This
integration generate an action  given  in terms  the Dirac
Born-Infeld action on the D2-brane wrapping on  $\Sigma $  in the
presence of the WZ term. This,  which involves also  the R-R 3-from
coupled to such a  D2-brane on $\Sigma $, produces a complex
quantity where its real part is identified with $S$.  In this
picture, small changes in $S$ produce large changes in $\Lambda
_{\sigma }$ for fixed ${\mu }$ by controlling fermionic zero mode
contributions. In this scenario, the mass behavior (\ref{eq17})
could be written using two parameters: the axion decay constant
$f_{\sigma }$, and the non-perturbative physics scale $\Lambda
_{\sigma }$ as follows,
\begin{equation}
m_{\sigma }\simeq {\mu }e^{-S}\frac{M_{s}}{f_{\sigma }}.
\end{equation}%
Thus stringy models are expected to possess a large number of axions, where
each axion is associated with a different modulus $S$. In connection with
superstring theory, one may consider models with various modulus. These can
be identified with R-R 3-form on 3-cycles of the internal Calabi-Yau
geometries. The corresponding number can be related to the number of the
complex deformations of such geometries. These stringy axions thus have a
mass spectrum spanning a vast number of orders of magnitude from the string
scale down to zero
\begin{equation}
0\lesssim m_{\sigma }\lesssim M_{s}.
\end{equation}%
This is the "\textit{string axiverse}" idea which corresponds to the
theoretical expectation stating that some large number of light and stable
axions should exist given the potential complexity of the internal geometry.

\section{Conclusion and open questions}

It has been highlighted that the axion fields are characterized by
two parameters: the axion decay constant and an energy scale
obtained from non-perturbative effect. This effect is associated
with instantons in superstring theory. Besides such energy scales,
it has been remarked that the number of such axions, in M-theory,
depends on the topology of the corresponding internal geometries.
Using string duality, this could be related to intersecting numbers
of 3-cycles in Calabi-Yau manifolds in the presence of D6-branes.

In this work, we have proposed a type IIA superstring quiver with an
open string axion field associated with the U(1)$_{PQ}$ global
symmetry. In the case of standard singularities, the Peccei-Quinn
realization extends the symmetry of the SM-Like and contributes to
the mediation of the electroweak symmetry breaking. The associated
open string axion scales have been investigated in terms of D2 and
D6-branes in type IIA compactifications. Using the fermion mass
scales as well as string theory data, we have first approached the
allowed axion window in terms of intersecting D6-branes wrapping
3-cycles belonging to the middle cohomology of the internal space.
Then we have discussed the non-perturbative effect on the open
string  axion field using a D2-brane wrapping a 3-cycle producing
1+3 dimensional instantons. According to the known data, we have
shown that the vacuum expectation value of the axion and the mass
scale of the involved new scalar provide a probe of the stringy
physics effect in the SM and a possible axionic DM candidate.

Our work comes up with many open questions. A natural question is
think about a mirror dual version in type IB superstring. In this
way, the D2-D6 brane systems could be replaced by D3-D7 ones using
the geometry of 4-cycles. Another question is to go beyond the
present study by introducing extra axion fields from D6-brane
models. This allows one to think about multi-instatons by
considering a set of D2-branes on 3-cycles producing non
perturbative string contributions. These could induce superpotential
coupling involving chiral charged matter fields given in terms of
fermionic zero modes living at the intersection of D6 and D2-branes.
We expect that such intersections bring new light on the axionic
dark matter discussions.  Alternatively, axions can be derived  from
closed string sectors. In superstring theory,  such fields can be
obtained either from the NS-NS or R-R forms wrapping appropriate
cycles of the internal space. In the case of type IIA superstring
theory, the closed string axion fields can be associated with the
R-R 3-form wrapping  3-cycles. These scalar fields can  carry charge
under the gauge fields living on the D6-brane wrapping on the same
3-cycles including the one associated  with the U(1)$_{PQ}$
symmetry.  In particular, the analogue of the  last term of the
lagrangian (\ref{eq4}) can be obtained from the Chern Simon actions
corresponding to the same D6-brane world-volume. It should be useful
to introduce  such fields in the ADM discussions by combing the open
and closed sectors  of string theory. We believe that this study
deserves a better understanding, and hope to report elsewhere on the
corresponding questions.
\bigskip

\textbf{Acknowledgements}: The authors would like to thank M. P. Garcia del
Moral for the collaboration on related topics.

\end{document}